\documentclass[12pt]{article}

\usepackage{mathrsfs}
\usepackage[T1]{fontenc}
\usepackage{mathpazo}
\usepackage{setspace}
\usepackage{amsfonts}
\usepackage{amssymb}
\usepackage{amsmath}
\usepackage{esint}                   
\usepackage{epsfig}
\usepackage{latexsym}
\usepackage{color}
\usepackage{graphicx}
\usepackage{hyperref}
\usepackage{nicefrac}
\usepackage[latin1]{inputenc}
\usepackage{pstricks}
\usepackage{slashed}
\usepackage{multirow}
\usepackage{ulem}
\usepackage{comment}
\usepackage[nosort]{cite}
\usepackage{datetime}
\usepackage{tikz-cd}
\usepackage{float}
\usepackage{capt-of}
\usetikzlibrary{decorations.markings}
\usepackage{caption}

\usepackage{tikz}
\usetikzlibrary{arrows,matrix,positioning,shapes,arrows}
\usetikzlibrary{shapes.geometric, arrows, calc, intersections}

\usepackage[greek,english]{babel}
\languageattribute{greek}{polutoniko}

\usepackage[vcentermath]{youngtab}

\usepackage{pifont}

\def\hybrid{\topmargin -30pt    \oddsidemargin 0pt 
        \headheight 0pt \headsep 0pt
        \textwidth 6.25in       
        \textheight 9.5in       
        \marginparwidth .875in
        \parskip 5pt plus 1pt   \jot = 1.5ex}

\hybrid

\def\baselinestretch{1.2}

\catcode`\@=11

\def\marginnote#1{}
\newcount\hour
\newcount\minute
\newtoks\amorpm
\hour=\time\divide\hour by60
\minute=\time{\multiply\hour by60 \global\advance\minute by-\hour}
\edef\standardtime{{\ifnum\hour<12 \global\amorpm={am}%
        \else\global\amorpm={pm}\advance\hour by-12 \fi
        \ifnum\hour=0 \hour=12 \fi
        \number\hour:\ifnum\minute<10 0\fi\number\minute\the\amorpm}}
\edef\militarytime{\number\hour:\ifnum\minute<10 0\fi\number\minute}

\def\draftlabel#1{{\@bsphack\if@filesw {\let\thepage\relax
   \xdef\@gtempa{\write\@auxout{\string
      \newlabel{#1}{{\@currentlabel}{\thepage}}}}}\@gtempa
   \if@nobreak \ifvmode\nobreak\fi\fi\fi\@esphack}
        \gdef\@eqnlabel{#1}}
\def\@eqnlabel{}
\def\@vacuum{}
\def\draftmarginnote#1{\marginpar{\raggedright\scriptsize\tt#1}}

\def\draft{\oddsidemargin -.5truein
        \def\@oddfoot{\sl preliminary draft \hfil
        \rm\thepage\hfil\sl\today\quad\militarytime}
        \let\@evenfoot\@oddfoot \overfullrule 3pt
        \let\label=\draftlabel
        \let\marginnote=\draftmarginnote
   \def\@eqnnum{(\theequation)\rlap{\kern\marginparsep\tt\@eqnlabel}%
\global\let\@eqnlabel\@vacuum}  }

\def\draft2{
        \def\@oddfoot{\sl preliminary draft \hfil
        \rm\thepage\hfil\sl\today\quad\militarytime}
        \let\@evenfoot\@oddfoot \overfullrule 3pt
        \let\marginnote=\draftmarginnote
   \def\@eqnnum{(\theequation)\rlap{\kern\marginparsep\tt\@eqnlabel}%
\global\let\@eqnlabel\@vacuum}  }

\def\preprint{\twocolumn\sloppy\flushbottom\parindent 2em
        \leftmargini 2em\leftmarginv .5em\leftmarginvi .5em
        \oddsidemargin -.5in    \evensidemargin -.5in
        \columnsep .4in \footheight 0pt
        \textwidth 10.in        \topmargin  -.4in
        \headheight 12pt \topskip .4in
        \textheight 6.9in \footskip 0pt
        \def\@oddhead{\thepage\hfil\addtocounter{page}{1}\thepage}
        \let\@evenhead\@oddhead \def\@oddfoot{} \def\@evenfoot{} }

\def\numberbysection{\@addtoreset{equation}{section}
        \def\theequation{\thesection.\arabic{equation}}}

\def\underline#1{\relax\ifmmode\@@underline#1\else
        $\@@underline{\hbox{#1}}$\relax\fi}

\def\titlepage{\@restonecolfalse\if@twocolumn\@restonecoltrue\onecolumn
     \else \newpage \fi \thispagestyle{empty}\c@page\z@
        \def\thefootnote{\fnsymbol{footnote}} }

\def\endtitlepage{\if@restonecol\twocolumn \else \newpage \fi
        \def\thefootnote{\arabic{footnote}}
        \setcounter{footnote}{0}}  

\catcode`@=12
\relax

\def\figcap{\section*{Figure Captions\markboth
        {FIGURECAPTIONS}{FIGURECAPTIONS}}\list
        {Figure \arabic{enumi}:\hfill}{\settowidth\labelwidth{Figure
999:}
        \leftmargin\labelwidth
        \advance\leftmargin\labelsep\usecounter{enumi}}}
 \relax
\def\tablecap{\section*{Table Captions\markboth
        {TABLECAPTIONS}{TABLECAPTIONS}}\list
        {Table \arabic{enumi}:\hfill}{\settowidth\labelwidth{Table
999:}
        \leftmargin\labelwidth
        \advance\leftmargin\labelsep\usecounter{enumi}}}
 \relax
\def\reflist{\section*{References\markboth
        {REFLIST}{REFLIST}}\list
        {[\arabic{enumi}]\hfill}{\settowidth\labelwidth{[999]}
        \leftmargin\labelwidth
        \advance\leftmargin\labelsep\usecounter{enumi}}}
 \relax
\makeatletter
\newcounter{pubctr}
\def\publist{\@ifnextchar[{\@publist}{\@@publist}}
\def\@publist[#1]{\list
        {[\arabic{pubctr}]\hfill}{\settowidth\labelwidth{[999]}
        \leftmargin\labelwidth
        \advance\leftmargin\labelsep
        \@nmbrlisttrue\def\@listctr{pubctr}
        \setcounter{pubctr}{#1}\addtocounter{pubctr}{-1}}}
\def\@@publist{\list
        {[\arabic{pubctr}]\hfill}{\settowidth\labelwidth{[999]}
        \leftmargin\labelwidth
        \advance\leftmargin\labelsep
        \@nmbrlisttrue\def\@listctr{pubctr}}}
 \relax
\makeatother

\def\be{\begin{equation}}
\def\ee{\end{equation}}
\def\ba{\begin{eqnarray}}
\def\ea{\end{eqnarray}}

\def\del{\partial}

\def\XXint#1#2#3{{\setbox0=\hbox{$#1{#2#3}{\int}$}
     \vcenter{\hbox{$#2#3$}}\kern-.5\wd0}}

\def\at{\tilde{\a}}
\def\bt{\tilde{\b}}
\def\gt{\tilde{\g}}

\def\r{\rho}
\def\a{\alpha}

\def\b{\beta}

\def\g{\gamma}
\def\G{\Gamma}
\def\d{\delta}
\def\D{\Delta}
\def\e{\epsilon}

\def\m{\mu}
\def\n{\nu}
\def\om{\omega}
\def\Om{\Omega}
\def\l{\lambda}

\def\s{\sigma}
\def\S{\Sigma}

\def\cH{{\cal H}}

\def\no{\noindent}

\def\qq{\qquad}

\def\IR{\relax{\rm I\kern-.18em R}}

\def\inv{^{\raise.0ex\hbox{${\scriptscriptstyle -}$}\kern-.05em 1}}

\def \ov {\over}

\begin{document}


\renewcommand{\theequation}{\thesection.\arabic{equation}}
\csname @addtoreset\endcsname{equation}{section}

\begin{titlepage}
\begin{center}

\renewcommand*{\thefootnote}{\arabic{footnote}}

\hfill HU-EP-23/57

\phantom{xx}
\vskip 0.5in

{\large \bf Supersymmetric backgrounds from $\lambda$-deformations}

\vskip 0.5in

{\bf Georgios Itsios}${}^{1a}$,\phantom{x} {\bf Konstantinos Sfetsos}${}^{2b}$\phantom{x}
and\phantom{x}{\bf Konstantinos Siampos}${}^{2c}$ \vskip 0.1in

${}^1$ Institut f\"{u}r Physik, Humboldt-Universit\"{a}t zu Berlin,\\
IRIS Geb\"{a}ude, Zum Gro{\ss}en Windkanal 2, 12489 Berlin, Germany\\

\vskip 0.11in

${}^2$ Department of Nuclear and Particle Physics, \\
Faculty of Physics, National and Kapodistrian University of Athens, \\
Athens 15784, Greece

\vskip .2in



\end{center}

\vskip .4in

\centerline{\bf Abstract}

\no
We provide the first supersymmetric embedding of an integrable $\lambda$-deformation to type-II supergravity. Specifically, that of the near horizon of the NS1-NS5 brane intersection, geometrically corresponding to $AdS_3\times S^3 \times T^4$. We show that the deformed background preserves $1/4$ of the maximal supersymmetry. In the Penrose limit we show that it preserves no-more than one half of the maximal supersymmetry.


\vfill
\no
 {\footnotesize
$^a$georgios.itsios@physik.hu-berlin.de \\
$^b$ksfetsos, $^c$konstantinos.siampos@phys.uoa.gr
}

\end{titlepage}
\vfill
\eject


\tableofcontents

\def\baselinestretch{1.2}
\baselineskip 20 pt

\newcommand{\eqn}[1]{(\ref{#1})}


\section{Introduction}

The near horizon limit of the NS1-NS5 brane-system~\cite{Duff:1995yh,Duff:1994an} corresponds geometrically to the space $AdS_3 \times S^3 \times T^4$. 
Together with appropriate three-form flux fields proportional to the volume forms of the $AdS_3$ and $S^3$ factors, it is a solution of the  common NS sector of type-IIA and type-IIB supergravities. This near horizon solution preserves half of the maximal supersymmetry  \cite{Duff:1993ye,Duff:1994an,Cowdall:1998bu}. 
 Moreover, each of the factors $AdS_3$ and $S^3$ with the flux fields corresponds to a current algebra exact conformal field theory (CFT) having a Lagrangian  description in terms of a WZW model \cite{Witten:1983ar} for the groups $SL(2,\IR)$ and $SU(2)$, respectively. A systematic procedure to deform WZW models for a general semi-simple group by breaking conformal invariance but preserving integrability has been devised in \cite{Sfetsos:2013wia} and is known as $\l$-deformation. This also includes the extensions to deformations based on more than one current algebra or even cosets CFTs \cite{Georgiou:2016zyo,Georgiou:2017jfi,Georgiou:2018hpd,Georgiou:2018gpe,Driezen:2019ykp,Hollowood:2014rla, Hollowood:2014qma,Sfetsos:2017sep}.

 A natural question is whether or not such deformations, when applied to the simplest case  of the $SL(2,\IR)\times SU(2)$ WZW model, can be promoted, with the inclusion of the $T^4$ factor, to a solution of type-II supergravity by turning on appropriate RR-fields. This will be then naturally called the  type-II supergravity embedding of the $\l$-deformation of the near horizon NS1-NS5 brane intersection background. Moreover, can such an embedding  be supersymmetric by preserving a fraction of the original half-supersymmetric near-horizon NS1-NS5 configuration? A supergravity embedding was indeed achieved in  \cite{Sfetsos:2014cea} albeit within the type-IIB* theory of \cite{Hull:1998vg} in which the RR-fields are purely imaginary which  is not satisfactory.\footnote{Besides this work type-III supergravity embeddings of $\l$-deformations based on (super) cosets have been constructed in \cite{Demulder:2015lva, Borsato:2016zcf, Itsios:2019izt, Chervonyi:2016ajp, Borsato:2016ose}. Such embeddings are not supersymmetric even in the limit of no deformation. Moreover, supergravity embeddings have been constructed for the closely related $\eta$-deformed models \cite{Klimcik:2002zj, Klimcik:2008eq,Delduc:2013fga,Delduc:2013qra}
 in \cite{Hoare:2015wia, Hoare:2015gda, Hoare:2018ngg, Lunin:2014tsa,Seibold:2019dvf}. The closest in spirit to our present work is the supersymmetric deformation of the $AdS_3 \times S^3 \times T^4$ solution studied in \cite{Hoare:2022asa}. This deformation also preserves one-quarter of the maximal supersymmetry and can be found by applying TsT transformations to the one-parameter Yang--Baxter integrable deformation~\cite{Klimcik:2002zj,Klimcik:2008eq,Delduc:2013fga}.}
In the present work, we manage to overcome this problem and provide the first such supersymmetric embedding in type-IIB supergravity, 
of the aforementioned $\l$-deformed near horizon NS1-NS5 brane intersection background.

The plan of the paper is as follows: In Section~\ref{TypeIISol}, we construct our $\nicefrac14$ supersymmetric type-IIB solution by turning on RR-fields. We check its supersymmetry and make contact with recent literature on one-quarter supersymmetric solutions with warped $AdS_2$ and $S^2$ factors. In Section ~\ref{Penrose}, we consider the Penrose limit around a null geodesic and the supersymmetry of the associated plane-wave solution. Details on the Killing spinor equations are contained in the Appendix~\ref{SupersymmetryVariations}.

\section{The supergravity solution}
\label{TypeIISol}

In this section we present the embedding of the $\l$-deformed model based on $SL(2,\mathbb{R}) \times SU(2)$ in type-IIB supergravity. 
A version of this was first found in \cite{Sfetsos:2014cea} but the embedding was in the type-IIB* theory of \cite{Hull:1998vg}.
However, in that work it was overlooked that the background  admits an analytic continuation of the coordinates mapping it to a solution of type-IIB supergravity with real RR-fields. In the notation of \cite{Sfetsos:2014cea} this reads\footnote{
 An alternative analytic continuation which results to a type-IIB solution is
\[
 \at \to \frac{\pi}{2} + i \at \, , \qquad \gt \to i \gt \, .
\]
In this paper we prefer to work with \eqref{AnalCont} as it is more appropriate for studying Penrose limits.
}
\begin{equation}
 \label{AnalCont}
 \at \to \frac{\pi}{2} + i \at \, , \qquad \bt \to i \frac{\pi}{2} + \bt \, .
\end{equation}
This analytic continuation does not change the signature of the spacetime and no change on the sign of the WZ level parameter $k$ is needed. 

\no
Alternative to the analytic continuation, one may use the general $\l$-deformed $\s$-model action \cite{Sfetsos:2013wia} with the group elements  for 
$SU(2)$ and $SL(2,\IR)$ given by
\be
\label{groops}
\begin{split}
& g_{SU(2)}=
\left(
\begin{array}{cc}
   \cos{\a} + i \sin\a \cos\b   & \sin\a \sin\b \ e^{-i\g}  \\
   -\sin\a \sin\b \ e^{i\g}   &   \cos{\a} - i \sin\a \cos\b 
\end{array}
\right)\ ,
\\
&  g_{SL(2,\IR)}= \left(
\begin{array}{cc}
-i\sinh{\tilde\a} -\cosh\tilde\a \sinh\tilde\b & -\cosh\tilde\a \cosh\tilde\b \ e^{-i\tilde\g} \\
\cosh\tilde\a \cosh\tilde\b \ e^{i\tilde\g} & -i\sinh{\tilde\a} +\cosh\tilde\a \sinh\tilde\b
\end{array}
\right)\ .
\end{split}
\ee
In the above, the $SU(2)$ group element is connected to the identity element reached for $\a=0$. Instead, the $SL(2,\IR)$ group element is not connected to the identity element for any real values of the angles $\at, \bt$ and $\gt$. Hence, we conclude that the identity is not part of the group manifold in the type-IIB supergravity solution we present below. This will be important when we will consider the non-Abelian T-duality (NATD) limit below. 

\begin{figure}[h!]
\centering
\begin{tikzpicture}
[
squarednode/.style={%
rectangle,
draw=black!60,
fill=white,
very thick,
minimum size=5mm,
text centered,
text width=5cm,
}
]
\node[squarednode] (NATD) {{\bf NATD} \\[5pt] $NATD(AdS_3 \times S^3) \times T^4$ \\[5pt] NS \ding{51}, RR \ding{51}};
\node[squarednode] (pureRR) [above=2.5cm of NATD] {{\bf Pure RR} \\[5pt] $AdS_3 \times S^3 \times T^4$ \\[5pt] NS \ding{56}, RR \ding{51}};
\node[squarednode] (pureNS) [left=2.5cm of pureRR] {{\bf Pure NS} \\[5pt] $AdS_3 \times S^3 \times T^4$ \\[5pt] NS \ding{51}, RR \ding{56}};

\draw[very thick, ->] (pureRR.south) -- node[anchor=west] {\,\, NATD} (NATD);
\draw[very thick, <->] (pureNS.south east) -- node [right,midway] {\,\, $\l$-def.}(NATD.north west);
\end{tikzpicture}
\captionsetup{width=.85\textwidth}
\caption{\small{Relation of the various solutions. The pure NS solution corresponds to $\l = 0$. When $\l$ approaches one, we find the NATD background.}}
\label{figure:solutionsWeb}
\end{figure}
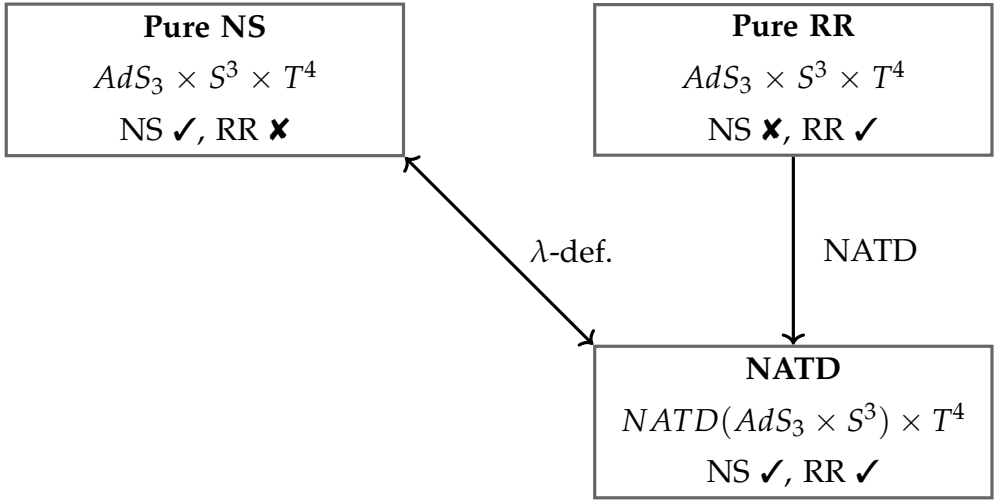

\no
The field content is summarized below and various interrelations are depicted in Figure~\ref{figure:solutionsWeb}.

\noindent\textbf{The metric:} The geometry in ten dimensions is the direct sum of the target space metric of the $\l$-deformed model on $SL(2,\mathbb{R}) \times SU(2)$ and a four-dimensional torus. The line element of this configuration is
\begin{equation}
 \label{Metric}
 \begin{aligned}
  ds^2 = & k \Big( \frac{1 + \l}{1 - \l} \, d\at^2 + \frac{1 - \l^2}{\tilde{\D}} \cosh^2\at \big( d\bt^2 - \cosh^2\bt \, d\gt^2 \big)
  \\[5pt]
  & + \frac{1 + \l}{1 - \l} \, d\a^2 + \frac{1 - \l^2}{\D} \sin^2\a \big( d\b^2 + \sin^2\b \, d\g^2 \big) \Big) 
  + dx^2_1 + dx^2_2 + dx^2_3 + dx^2_4 \, ,
 \end{aligned}
\end{equation}
where
\begin{equation}
 \begin{aligned}
  & \tilde{\D} = (1 + \l)^2 \cosh^2 \tilde \a - (1 - \l)^2 \sinh^2 \tilde \a \, ,
  \\[5pt]
  & \D = (1 - \l)^2 \, \cos^2 \a + (1 + \l)^2 \, \sin^2 \a \, .
 \end{aligned}
\ee
The first line in \eqref{Metric} corresponds to the target space metric of the $\l$-deformed model on $SL(2,R)$, while the part spanned by $(\a, \b, \g)$ to that of the $SU(2)$ $\l$-model. The four-torus is parametrized by the coordinates $x_i \, (i = 1, \ldots , 4)$. 

\no
Notice that when $\l = 0$ the six-dimensional space transverse to the torus is just $AdS_3 \times S^3$. Although the presence of the deformation breaks the isometry of $AdS_3 \times S^3$, a subspace with topology $AdS_2 \times S^2$ is still present. When $\at$ approaches $\pm \infty$ the subspace parametrized by $(\at , \bt , \gt)$ looks like $\mathbb{R} \times AdS_2$. On the other hand the topology of the subspace parametrized by $(\a , \b , \g)$ looks different in the neighborhood of the points $\a = 0, \pi$ and $\a = \nicefrac{\pi}{2}$. Specifically, near $\a = 0, \pi$ it looks like $\mathbb{R}^3$, while near $\a = \nicefrac{\pi}{2}$ it behaves like $\mathbb{R} \times S^2$.

\no
For later convenience we also introduce the orthogonal frame
\begin{equation}
 \label{Frame}
 \begin{aligned}
  & e^0 = \sqrt{k \frac{1 - \l^2}{\tilde{\D}}} \cosh\at \, \cosh\bt \, d\gt \, , \quad
     e^1 = \sqrt{k \frac{1 - \l^2}{\tilde{\D}}} \cosh\at \, d\bt \, , \quad
     e^2 = \sqrt{k \frac{1 + \l}{1 - \l}} d\at \, ,
  \\[5pt]
  & e^3 = \sqrt{k \frac{1 + \l}{1 - \l}} d\a \, , \quad
     e^4 = \sqrt{k \frac{1 - \l^2}{\D}} \sin\a \, d\b \, , \quad
     e^5 = \sqrt{k \frac{1 - \l^2}{\D}} \sin\a \, \sin\b \, d\g \, ,
  \\[5pt]
  & e^6 = dx_1 \, , \qquad e^7 = dx_2 \, , \qquad e^8 = dx_3 \, , \qquad e^9 = dx_4 \, .
 \end{aligned}
\end{equation}
As usual this is associated to the ten-dimensional tangent frame metric on $\mathbb{R}^{1,9}$
\begin{equation}
\label{metric.frame}
ds^2=\eta_{ab}e^ae^b\,,\quad g_{\mu\nu}=\eta_{ab}e^a{}_\mu e^b{}_\nu\,,\quad e^a=e^a{}_\mu dx^\mu\,,\quad \eta=\text{diag}(-1,1,\cdots,1)\,,
\end{equation}
where Greek indices denote the curved ones, while Latin ones those of the tangent space.

\noindent\textbf{The dilaton:} The NS sector contains also a non-trivial dilaton arising from integrating out the gauge fields in the construction \cite{Sfetsos:2014cea}. This takes the simple form 
\begin{equation}
 \label{Dilaton}
 e^{- 2 \Phi} = {\D \, \tilde{\D}\ov  (1+\l)^4} \, .
\end{equation}
The inclusion of the $\l$-depended factor is arbitrary and affects the form of the RR-flux fields below by a related overall factor. We have chosen this 
factor in such a way that all the background fields are invariant under the non-perturbative symmetry
\be
\label{syyym}
\l\to {1\ov \l} \ ,\qq k\to -k\ ,\qq \a\to -\a\ ,\qq \at\to -\at \ ,
\ee
found in the general contest of the $\l$-deformed models in \cite{Itsios:2014lca}. This symmetry is satisfied by construction by the metric \eqn{Metric} and the NS two-form \eqn{NS2form} below.

\noindent\textbf{The NS two-form:} The NS two-form has contribution both in the $SL(2, \mathbb{R})_\l$ and in the $SU(2)_\l$ directions, namely
\begin{equation}
 \label{NS2form}
 \begin{split}
 & B_2  = k \Big( - \a + \frac{(1 - \l)^2}{\D} \cos\a \, \sin\a \Big) \sin\b \, d\b \wedge d\g
 \\
&  \qquad  +  k \Big( \at + \frac{(1 - \l)^2}{\tilde{\D}} \cosh\at \, \sinh\at \Big) \cosh\bt \, d\bt \wedge d\gt \, .
 \end{split}
\end{equation}

\noindent\textbf{The RR-forms:} This solution has all the RR forms turned on when $\l \ne 0$. These are better expressed in terms of the frame \eqref{Frame} as
\begin{equation}
 \label{RRforms}
 \begin{aligned}
  & F_1 = - \m {1 - \l\ov 1+\l} \big( \cos\a \cosh\at \, e^2 - \sinh\at \sin\a \, e^3 \big) \, ,
  \\[5pt]
  & F_3 = \m \bigg( {\Big({1 - \l\ov 1+\l}\Big)^2} \cos\a \sinh\at \big( e^{012} + e^{345} \big) + \sin\a \cosh\at \big( e^{245} - e^{013} \big) \bigg) \, ,
  \\[5pt]
  & F_5 = - \m {1 - \l\ov 1+\l} (1 + \star) \big( \sin\a \sinh\at \, e^{01245} + \cos\a \cosh\at \, e^{01345} \big) \, ,
 \end{aligned}
\end{equation}
where $F_5$ is self-dual\footnote{Our conventions for the Hodge dual on a p-form in a D-dimensional spacetime are that
\begin{equation*}
(\star F_p)_{\mu_{p+1}\ldots\mu_D}=\frac{1}{p!}\sqrt{|g|}\varepsilon_{\mu_1\ldots\mu_D}F_p^{\mu_1\ldots\mu_p}\,,
\end{equation*}
where $\varepsilon_{0\ldots9}=1$. From the above we find $\star\star F_p=s(-1)^{p(D-p)}F_p$, where $s$ is the signature of the spacetime which in our case is taken to be mostly plus, see Eq.~\eqref{metric.frame}.\label{Hodge}} while $\m$ depends on $\l$ and $k$ through
\begin{equation}
\label{mmm}
 \m = \frac{4 \l}{\sqrt{k (1 - \l)(1+\l)^3}} \, .
\end{equation}

\no
It is interesting to point out that the background described above interpolates between the type-IIB solution on $AdS_3 \times S^3 \times T^4$ with only NS fields, and the type-IIB solution arising from a non-Abelian T-duality (NATD) transformation on $AdS_3 \times S^3 \times T^4$ with a RR three-form and vanishing NS form. The former corresponds to $\l = 0$ while the latter can be seen as a correlated limit $\l \rightarrow 1, \, k \rightarrow \infty$ combined with a zoom-in along $\at$ and $\a$. Moreover, as we will see later, when $\l = 0$ the supergravity solution preserves $16$ supercharges while for $\l \neq 0$ supersymmetry is broken to $8$ supercharges. This is in agreement with the fact that the deformation breaks the isometries of the background.

\subsection{The non-Abelian T-dual limit}

It is known that when $\l$ approaches the identity, the $\l$-deformed model becomes singular. It order to make sense of the geometry a correlated limit in which $k$ is also taken to infinity has to be taken. Then, the result on a group reduces to the non-Abelian T-dual of the corresponding Principal Chiral Model (PCM) as long as, at the same time, a zoom-in near the identity element of the group, by rescaling the group parameters appropriately with inverse powers of $k$, is also taken \cite{Sfetsos:2013wia}.

\no
In turns that the background \eqn{Metric}, \eqn{Dilaton} and \eqn{RRforms} does not admit a non-Abelian T-dual limit. The reason is that,  
as noted before, there are no  real value for the angles $\at, \bt$ and $\gt$ for which the identity element of $SL(2,\IR)$ group element  in 
\eqn{groops} can be reached. 
Nevertheless, at the supergravity level described in this section, the NATD limit can be taken by setting
\begin{equation}
\label{NATDlimit}
\at = i \frac{\pi}{2} + \frac{\r}{2 k} \, , \qquad \a = \frac{r}{2 k} \, , \qquad \l = 1 - \frac{1}{k}
\end{equation}
and then sending $k$ to infinity.\footnote{\label{foot3}Curiously,  a coordinate transformation of the form $ \at \to i \frac{\pi}{2} + \at$ leads to a solution of type-IIB$^*$ for all values of $\l$ except for the non-Abelian T-dual limit \eqref{NATDlimit} as it is explained below.} 
This analytic continuation can be easily seen to effectively connect the $SL(2,\IR)$ group element in \eqn{groops} to the identity which is a necessary condition for the non-Abelian T-duality limit to 
exist.
Moreover, in the NATD limit the dilaton diverges with an imaginary constant. Shifting the dilaton with a $k$-dependent 
 constant absorbs this divergence. In doing so, one has to also rescale the RR-forms by an appropriate imaginary $k$-dependent constant. 
 This guarantees that the NATD limit solves the supergravity equations of motion and eventually provides a real solution of type-IIB supergravity with the 
 various fields presented below. 
 
\noindent\textbf{The NS sector:} After the NATD limit \eqref{NATDlimit} the metric, the dilaton and the NS two-form read
\begin{equation}
\label{NSNATD}
\begin{aligned}
ds^2 & = \frac{1}{2} \Big( d\r^2 + \frac{\r^2}{\r^2 - 1} \big( d\bt^2 - \cosh^2\bt d\gt^2 \big) + dr^2 + \frac{r^2}{r^2 + 1} \big( d\b^2 + \sin^2\b d\g^2 \big) \Big)
\\[5pt]
&\quad  + dx^2_1 + dx^2_2 + dx^2_3 + dx^3_4 \ , \qquad
e^{-2 \Phi} = \big( \r^2 - 1 \big) \big( r^2 + 1 \big) \ ,
\\[5pt]
B_2 & = \frac{1}{2} \Big( \frac{\r^3}{\r^2 - 1} \cosh\bt d\bt \wedge d\gt - \frac{r^3}{r^2 + 1} \sin\b d\b \wedge d\g \Big) \, .
\end{aligned}
\end{equation}
Notice that in order for the metric to preserve the correct signature, the coordinate $\r$ must take values $|\r| > 1$. 
For later convenience in comparing  with the literature, we have  multiplied the resulting from the limit expression for $e^{-2 \Phi}$  by a factor of $16$. That affects the expressions for the RR-flux fields below which, in comparison with the results one gets from the limit, 
have been multiplied by a factor of $4$.

\noindent\textbf{The RR sector:} Taking into account the rescaling of the RR forms due to the shift of the dilaton, the limit \eqref{NATDlimit} results to
\begin{equation}
\label{RRNATD}
\begin{aligned}
F_1 & = \r d\r - r dr \, ,
\\[5pt]
F_3 & = \frac{1}{2} \Bigg( \frac{\r^2}{\r^2 - 1} \cosh\bt \big( d\r - \r r dr \big) \wedge d\bt \wedge d\gt - \frac{r^2}{r^2 + 1} \sin\b \big( dr + r \r d\r \big) \wedge d\b \wedge d\g \Bigg) \, ,
\\[5pt]
F_5 & = - \frac{1}{4} \frac{\r^2 r^2}{\big( \r^2 - 1 \big) \big( r^2 + 1 \big)} \cosh\bt \sin\b (1 + \star) \big( \r dr + r d \r \big) \wedge d\bt \wedge d\gt \wedge d\b \wedge d\g \, .
\end{aligned}
\end{equation}
It is worth to mention that the NATD solution falls in the class of $AdS_2 \times S^2 \times \text{CY}_2$ backgrounds with $8$ supercharges constructed in \cite{Lozano:2021rmk}. One can see this explicitly by setting
$
\hat{h}_4 = h_8 = \frac{u}{2} $ and $u = - r \ , 
$
in equations $(4.6)$ and $(4.7)$ of that paper (note the difference in notation $(r , \r)_{\text{there}} = (\r , r)_{\text{here}}$).

\subsection{Supersymmetry}

We turn next to the analysis of the supersymmetry for the solution given in \eqref{Metric}, \eqref{Dilaton}, \eqref{NS2form} and \eqref{RRforms}. Due to the fact that the RR fields vanish when $\l = 0$ it is instructive to distinguish this case from the $\l \ne 0$ one.

\noindent\textbf{The $\l = 0$ case:} As it is explained in appendix \ref{SUSYanalysis} when $\l = 0$ the dilatino equation is solved by imposing a single projection, namely the one given in \eqref{ProjectorLambda0}. Also, the gravitino equations can be easily integrated resulting to the Killing spinor\footnote{
We have simplified the expression omitting the tensor product. More precisely, we assume
\begin{equation*}
 \s_i \to \s_i \otimes \mathbb{1}_{32} \quad (i = 1,2,3) \, , \qquad \G^a \to \mathbb{1}_2\otimes \G^a \quad (a = 0 , \ldots , 9) \, .
\end{equation*}
\label{Gamma.sigma}
}
\begin{equation}
 \begin{aligned}
  \e = & \exp \Big( - \frac{\at}{2}\s_3 \G^{01} \Big) 
 \exp \Big( \frac{\bt}{2} \s_3\G^{02} \Big) 
 \exp \Big( - \frac{\gt}{2} \s_3\G^{12} \Big)  
 \\[5pt]
 &  \exp \Big( - \frac{\a}{2} \s_3\G^{45} \Big) 
 \exp \Big( \frac{\b}{2} \G^{34} \Big) 
 \exp \Big( \frac{\g}{2} \G^{45} \Big)  \eta \, ,
 \end{aligned}
\end{equation}
where the indices in the Gamma-matrices are in accordance with the frames defined in \eqn{Frame} and  $\eta$ is a constant spinor satisfying the projection
\begin{equation}
 \label{ProjectorLambda0eta}
 \G^{0 \cdots 5} \eta = - \eta \, .
\end{equation}
The latter means that the type-IIB solution on $AdS_3 \times S^3 \times T^4$ with trivial RR sector preserves $16$ supercharges, already known in the literature \cite{Duff:1993ye}.

\noindent\textbf{The $\l \ne 0$ case:} The presence of the RR fields when $\l \ne 0$ requires that we impose an extra projection in order for the dilatino equation to vanish. Indeed, together with \eqref{ProjectorLambda0} one has also to consider \eqref{ProjectorLambdane0}. Moreover, after imposing the aforementioned projections and the chirality condition of the type-IIB Majorana--Weyl spinors one can integrate the gravitini. Doing so, leads to an expression for the Killing spinor which now depends non-trivially on $\l$
\begin{equation}
 \label{KillingSpinorLambda}
\begin{aligned}
  \e = & \exp \left( - \frac{1}{2} \tanh^{-1} \Big( \frac{1 - \l}{1 + \l} \tanh\at \Big)  \s_3\G^{01}\right)
 \exp \left( \frac{\bt}{2}\s_3 \G^{02}  \right) 
 \exp \left( - \frac{\gt}{2}\s_3 \G^{12} \right) 
 \\[5pt]
 &  \exp \left( - \frac{1}{2} \tan^{-1} \Big( \frac{1 + \l}{1 - \l} \tan\a \Big)\s_3 \G^{45} \right) 
 \exp \left( \frac{\b}{2} \G^{34}\right) 
 \exp \left( \frac{\g}{2} \G^{45} \right)  \eta \, .
 \end{aligned}
\end{equation}
Again, $\eta$ is a constant spinor which now satisfies
\begin{equation}
 \label{ProjectorLambdane0eta}
 \s_1\G^{01} \eta = - \eta\ ,
\end{equation}
together with the independent projector \eqref{ProjectorLambda0eta}. The fact that one is forced to impose the second projection \eqref{ProjectorLambdane0eta} when $\l \ne 0$ suggests that the deformed solution preserves $8$ supercharges. Therefore the deformation 
breaks the original supersymmetry by half. 

\paragraph{Supersymmetry of the NATD:}
In the case of the NATD solution given in \eqref{NSNATD} and \eqref{RRNATD} one can analyse the dilatino and gravitino equations from scratch or simply take the limit \eqref{NATDlimit} in \eqref{KillingSpinorLambda} and \eqref{ProjectorLambdane0}. Either way one finds the Killing spinor
\begin{equation}
 \begin{aligned}
  \e = & \exp \Big( - \frac{1}{2} \s_3\G^{01} \coth^{-1} \r \Big) 
 \exp \Big( \frac{\bt}{2} \s_3\G^{02} \Big) 
 \exp \Big( -\frac{\gt}{2} \s_3\G^{12} \Big)  
 \\[5pt]
 &  \exp \Big( - \frac{1}{2} \s_3 \G^{45}\tan^{-1} r  \Big) 
 \exp \Big( \frac{\b}{2} \G^{34} \Big) 
 \exp \Big( \frac{\g}{2} \G^{45} \Big)  \eta \, ,
 \end{aligned}
\end{equation}
with $\eta$ being a constant spinor satisfying \eqref{ProjectorLambda0eta} and \eqref{ProjectorLambdane0eta}. We see that the NATD solution also preserves $8$ supercharges.

It is worth making the connection with a new type-IIB class of solutions on a warped product of $AdS_2 \times S^2 \times CY_2 \times \S_2$ preserving $8$ supercharges was constructed recently in \cite{Legramandi:2023fjr}. This class generalizes the results of \cite{Lozano:2021rmk} and is governed by a system of partial differential equations that is reminiscent of D3-D7-brane configurations and a harmonic function on $\S_2$. The NS sector of these solutions is given in eq. (5.35) of \cite{Legramandi:2023fjr}, while the RR fields can be obtained by combining eq. (5.36) and (2.1). It turns out that the solution discussed here, namely the one given in \eqref{Metric}, \eqref{Dilaton}, \eqref{NS2form} and \eqref{RRforms}, fits in the class mentioned above. This is in agreement with the outcomes from the supersymmetry analysis that we present later. For the moment we provide the relation between our solution and the class of \cite{Legramandi:2023fjr}
\begin{equation}
 \begin{aligned}
  & y_1 = \a \, , \quad y_2 = \at \, , \quad h_3 = h_7 = \frac{1 + \l}{1 - \l} \, u \, , \quad u = \cosh\at \, \sin\a \, , \quad c_0 = \frac{1 - \l}{1 + \l} \, ,
  \\[5pt]
  & X_1^{(1,1)} = X_2^{(1,1)} = 0 \, .
 \end{aligned}
\end{equation}
To avoid confusion, we denote the coordinates on $\S_2$ by $y_1 \, , y_2$ instead of $x_1 \, , x_2$ that is used in \cite{Legramandi:2023fjr}. Also, for simplicity we take $k = 1$. With the above identifications we see that the metric \eqref{Metric} and the dilaton \eqref{Dilaton} match those in eq. (5.35) of \cite{Legramandi:2023fjr}. However, for the NS three-form we find a difference by an overall minus sign. The same happens with the RR three-forms. Therefore, no problem occurs in view of the symmetry $(H_3 \, , F_3) \leftrightarrow (- H_3 \, , - F_3)$ of the type-IIB equations of motion. For the RR one- and three-forms we also find precise agreement.  Some discrepancies for the five-form  are  explained by the fact that in our case $F_5$ is self-dual while in \cite{Legramandi:2023fjr} anti-self-dual (in our conventions for Hodge duality, as described in footnote~\ref{Hodge}).

\section{Penrose limit}
\label{Penrose}

We continue with the study of null geodesics in the geometry of the $\l$-deformed background and the derivation of the corresponding plane-wave solutions through a Penrose limit  \cite{PenroseLim}.


\subsection{Null geodesics}

In order to find null geodesics in the geometry of Section \ref{TypeIISol} we consider a particle that is moving along the $U(1)$ isometries realized by shifts in the coordinates $\gt$ and $\g$. In other words the velocity of the particle has non-vanishing components only along those directions. Requiring that there is no acceleration, i.e. $\ddot \g=\ddot \gt=0$, the equations of motion of the particle reduce to
\begin{equation}
 c^2_{\gt} \, \partial_\m g_{\gt\gt} + c^2_{\g} \, \partial_\m g_{\g\g} = 0 \, ,
\end{equation}
where $c_{\gt}=\dot \gt$ and $c_{\g}=\dot \g$ are the velocities along $\gt$ and $\g$, respectively, and the index $\m$ labels the directions of the ten-dimensional spacetime. To arrive to the above result, we took into account that the metric is diagonal and that it does not depend on $\gt$ and $\g$. This condition implies that
\begin{equation}
 \frac{\sinh(2\at) \cosh^2\bt}{\tilde{\D}(\at)^2} = \frac{\cosh^2\at \sinh(2 \bt)}{\tilde{\D}(\at)} = \frac{\sin(2\a) \sin^2\b}{\D(\a)^2} = \frac{\sin^2\a \sin(2 \b)}{\D(\a)} = 0 \, .
\end{equation}
Having in mind that the $U(1)$ directions $\gt$ and $\g$ should not shrink to zero size prevents us from considering $\a = \b = 0$. Thus the only solution of the equations of motion above that makes sense is
\begin{equation}
 \label{Geodesic}
 \at = \bt = 0 \, , \qquad \a = \b = \frac{\pi}{2} \, .
\end{equation}
The null condition simply translates to considering equal velocities along the directions $\gt$ and $\g$, i.e. $c_{\gt} = c_\g$.

\subsection{Plane wave solution}
\label{section.pp}

Here we derive the plane wave solution that arises by applying the Penrose limit around the geodesic \eqref{Geodesic} of the $\l$-deformed background presented in section \ref{TypeIISol}. For this reason we set
\begin{equation}
 \begin{aligned}
  & \at = \sqrt{\frac{1 - \l}{1 + \l}} \, \frac{z_1}{\sqrt{k}} \, , \qquad
 \a = \frac{\pi}{2} + \sqrt{\frac{1 - \l}{1 + \l}} \, \frac{z_2}{\sqrt{k}} \, ,
 \\
 & \bt = \sqrt{\frac{1 + \l}{1 - \l}} \, \frac{z_3}{\sqrt{k}} \, , \qquad
 \b = \frac{\pi}{2} + \sqrt{\frac{1 + \l}{1 - \l}} \, \frac{z_4}{\sqrt{k}} \, ,
 \\
 & \gt = u \, , \qquad\qquad\qquad\,\,\,\,\, \g = u + \frac{1 + \l}{1 - \l} \, \frac{v}{k} \, .
 \end{aligned}
\end{equation}
The fact that the coefficients in front of $u$ for $\gt$ and $\g$ are equal ensures that the geodesic is null. After applying this to the supergravity fields of the solution and considering large values for $k$ the background drastically simplifies to
\begin{equation}
 \label{ppWaveBackground}
 \begin{aligned}
  & ds^2 = 2 du dv + d\vec{x}^2_4 + d\vec{z}^2_4  -  \bigg[\Big( \frac{1 - \l}{1 + \l} \Big)^4 \big( z^2_1 + z^2_2 \big) +z^2_3 + z^2_4 \bigg]
    \, du^2 \, , 
  \\[5pt]
  & H_3 = 2 \frac{1 + \l^2}{( 1 + \l )^2} du \wedge \big( dz_1 \wedge dz_3 - dz_2 \wedge dz_4 \big) \  , \qquad \Phi =0 \, ,
  \\[5pt]
  & F_3 = 4 \, {\l\ov (1+\l)^2} \, du \wedge \big( dz_1 \wedge dz_4 + dz_2 \wedge dz_3 \big) \  .
 \end{aligned}
\end{equation}
where we note that $F_1$ and $F_5$ in \eqn{RRforms} have not survived the Penrose limit. The above content provides a solution of the type-IIB supergravity on a pp-wave geometry. Notice that the mass terms (coefficient of $du^2$ in the line element) break the $O(8)$ global symmetry in the transverse directions to $O(4) \times O(2) \times O(2)$ when $\l \ne 0$. For $\l = 0$ the symmetry of the metric enhances to $O(4) \times O(4)$ and for $\l = 1$ to $O(6) \times O(2)$. Also note that \eqn{ppWaveBackground} is invariant under $\l\to\nicefrac1\l$ which is the left over of the non-perturbative symmetry \eqn{syyym} in the Penrose limit.

\subsection{Supersymmetry} 

At this point we would like to inquire whether the pp-wave solution found above admits supernumerary supercharges. We already know that when $\l = 0$, supersymmetry on the pp-wave enhances from $16$ (which is the minimum) supercharges to $24$ \cite{Sadri:2003ib}. The question is what happens when $\l \ne 0$. This can be inferred by examining the dilatino and gravitino variations for the background \eqref{ppWaveBackground}. Below we provide a summary of our findings, and we leave the details in Appendix \ref{ppsusy}.

The Killing spinor that solves the supersymmetry variations depends on $\l$ and takes the form
\begin{equation}
\label{spinorexpression}
 \e = (1+ \G^{-} \Om) \chi(u) \, .
\end{equation}
The $\l$ dependence is hidden in $\Om$ and $\chi(u)$. Here $\Om$ is a matrix linear in the coordinates $z_i , \, x_i \, (i = 1, \ldots ,4)$ which can be represented in terms of the $\G$'s as in \eqref{Omega.matrix}. Recall that we work in the frame \eqref{ppFrame}. Additionally, $\chi(u)$ is expressed in relation to a constant spinor $\eta$ as shown below
\begin{equation}
\label{spinor.sol}
\chi(u)=\exp\left(\frac{u}{2}\frac{1+\l^2}{(1+\l)^2}\s_3 \big(\G^{13} - \G^{24}\big)+
 \frac{u}{2}\frac{\l}{(1+\l)^2}\s_1\G^{-+}\big(\G^{14}+\G^{23}\big)\right)\eta\,.
\end{equation}
Furthermore, $\chi(u)$ and $\eta$ are constrained by \eqref{pp.conditions} and \eqref{dilatinochi}.

Looking at \eqref{pp.conditions} and \eqref{dilatinochi} we can distinguish the cases $\l = 0$ and $\l \ne 0$. Obviously, when $\l = 0$ \eqref{pp.conditions} are trivially satisfied, while \eqref{dilatinochi} becomes
\begin{equation}
 \G^{-} \big( 1 - \G^{1234} \big) \eta = 0 \, .
\end{equation}
This amounts to $8$ supernumerary supercharges or $24$ in total. On the other hand, when $\l \ne 0$, the conditions \eqref{pp.conditions} are satisfied only when
\begin{equation}
 \G^{-} \eta = 0 \, .
\end{equation}
This automatically solves \eqref{dilatinochi}. Consequently, in this case there are no supernumerary supercharges and the pp-wave background preserves the minimal supersymmetry, i.e. $16$ supercharges.

\section{Conclusions}

We promoted the integrable $\l$-deformed model based on $SL(2, \mathbb{R}) \times SU(2)$ to a solution of the type-IIB supergravity by including the necessary RR fluxes. This solution utilizes a group element disconnected from the identity and can be obtained as an analytic continuation of the type-IIB$^*$ solution found in~\cite{Sfetsos:2014cea}. In the limit where the deformation parameter is zero, we recover the geometry of $AdS_3 \times S^3 \times T^4$ supported only by a NS three-form. This geometry originates from the near-horizon limit of the NS1-NS5 brane setup and preserves $16$ supercharges. Although in the presence of the deformation, the full group of isometries of $AdS_3 \times S^3$ is broken, the geometry has a manifest $AdS_2 \times S^2$ topology. We showed that supersymmetry is reduced by half, and in fact the ten-dimensional background fits into the class of solutions recently found in \cite{Legramandi:2023fjr}. This is the first known example of a supersymmetric $\l$-deformed supergravity solution. Finally, we consider the Penrose limit along a null geodesic in the deformed spacetime. This results in a type-IIB solution on a plane-wave geometry that captures a dependence on the deformation parameter. In the absence of deformation, there exist $8$ supernumerary supercharges  \cite{Sadri:2003ib}, while for non-zero deformation, the pp-wave background preserves the minimum amount of supersymmetry of $16$ supercharges. It turns out that this is not the only supersymmetric embedding of integrable $\l$-deformations to supergravity, since along similar lines, the embedding of the $\l$-deformed $AdS_3 \times S^3 \times S^3 \times S^1$ can also be constructed \cite{Itsios:2023uae}.

The aforementioned features leave room for further investigations of the deformed background. It is interesting to consider whether the deformed solution admits supersymmetric embeddings of probe branes \cite{Cederwall:1996ri,Bergshoeff:1996tu,Arean:2004mm} by analyzing the $\kappa$-symmetry condition. Additionally, we have already observed that the deformed geometry maintains an $AdS_2$ subspace. Therefore, it is natural to explore the holographic dual of our construction and study the thermal effects by constructing the corresponding  black hole solution, which amounts to turning on temperature in the holographic dual system. Finally, it would be worth checking whether our supersymmetric $\l$-deformed $AdS_3\times S^3\times T^4$ background can be derived from the integrable $\l$-deformed supercoset $\s$-model constructed in~\cite{Hoare:2022vnw}, also argued  to preserve 8 supercharges.

\section*{Acknowledgements}

We thank Niall Macpherson for a very useful correspondence.\\
The research work of G.~Itsios  is supported by the Einstein Stiftung Berlin via the Einstein International Postdoctoral Fellowship program 
 ``Generalised dualities and their holographic applications to condensed matter physics'' (project number IPF-2020-604). G.~Itsios is also supported by the Deutsche Forschungsgemeinschaft (DFG, German Research Foundation) via the Emmy Noether program ``Exploring the landscape of string theory flux vacua using exceptional field theory'' (project number 426510644).\\
The research work of K.~Sfetsos and K.~Siampos was supported by the Hellenic Foundation for Research and Innovation (H.F.R.I.) under the ``First Call for H.F.R.I. Research Projects to support Faculty members and Researchers and the procurement of high-cost research equipment grant'' (MIS 1857, Project Number: 16519).

\appendix

\section{Details on supersymmetry}
\label{SupersymmetryVariations}

In this appendix we include the conventions for the supersymmetry variations of the dilatino and gravitino in type-IIB supergravity and also their explicit expressions for the background of the Section \ref{TypeIISol}.



\subsection{Conventions}

Let us introduce our conventions for the dilatino\footnote{The dilatino and deformation parameter are denoted by the same letter $\l$, but this should not be a source of confusion.} and gravitino supersymmetry variations \cite{Hassan:1999bv}
\begin{equation} 
\label{SUSYvariations}
 \begin{aligned}
  & \d\l= \frac{1}{2} \slashed{\partial} \Phi \e - \frac{1}{24} \s_3\slashed{H} \e + \frac{e^{\Phi}}{2} \Big[ \big( i \s_2 \big) \slashed{F}_1 + \frac{1}{12} \s_1\slashed{F}_3  \Big] \e \, ,
  \\[5pt]
  & \d\psi_\m = D_\m \e - \frac{1}{8} H_{\m\n\r} \s_3\G^{\n\r}  \e - \frac{e^\Phi}{8} \Big[\big( i \s_2 \big) \slashed{F}_1  + \frac{1}{6} \s_1\slashed{F}_3  + \frac{1}{2 \cdot 5!} \big( i \s_2 \big)\slashed{F}_5  \Big] \G_\m \e \, ,
 \end{aligned}
\end{equation}
where $(i\s_2)$ is kept in parenthesis to illustrate that it is real and the slash means contraction of the spacetime indices with antisymmetric products of $\G$-matrices. In particular, we take $\slashed{\partial} = \G^\m \partial_\m=e^\m{}_a\G^a\del_\mu$ and $\slashed{A}_n = A_{\m_1 \ldots \m_n} \G^{\m_1 \ldots \m_n}$. Also, we have abbreviated tensor products of Gamma-matrices according to footnote \ref{Gamma.sigma}. Moreover, $\s_i \, (i = 1, 2, 3)$ are the Pauli matrices, $\e$ represents a doublet
\begin{equation}
 \e =
 \begin{pmatrix}
  \e_+
  \\
  \e_-
 \end{pmatrix}\ ,
\end{equation}
of two Majorana--Weyl spinors and satisfies the chirality condition 
\be
\label{chiral}
\G^{0 \cdots 9} \e = - \e\ ,
\ee 
in the tangent frame, while $F_5$ is self-dual (as defined in footnote~\ref{Hodge}). It is understood that the Pauli matrices have a $GL(2)$ action on the doublet and for convenience we suppress the $GL(2)$ indices. Finally, $D_\m$ is the spinor covariant derivative
\begin{equation}
 D_\m = \partial_\m + \frac{1}{4} \slashed{\om}_\m \, ,
\end{equation}
with $\om_{\m ab}=e^c{}_\m \om_{cab}$ being the spin-connection, which is antisymmetric in the last two indices for the tangent frame metric.

\subsection{Supersymmetry variations for the $\l$-deformed solution}
\label{SUSYanalysis}

We move to the analysis of the supersymmetry for the background of the Section \ref{TypeIISol}. Starting with the dilatino equation \eqref{SUSYvariations}, this greatly simplifies by imposing the projection in the tangent frame
\begin{equation}
 \label{ProjectorLambda0}
 \G^{0 \ldots 5} \e = - \e \, .
\end{equation}
It can be easily checked that this solves the dilatino equation \eqref{SUSYvariations} for $\l = 0$. However, for generic $\l \ne 0$, after using \eqref{ProjectorLambda0}  we arrive at

\begin{equation}
\label{fhgu1}
 \begin{aligned}
   \d\l & = - \frac{\m}{2} \left( \frac{(1 - \l)^2 \sinh^2\at}{\tilde{\D}} \s_3\G^{012} 
 - \frac{(1 - \l)^2 \cos^2\a}{\D}\s_3 \G^{345}\right.
 \\[5pt]
 & + \frac{(1 - \l^2) \sin\a \cos\a}{\D} \G^3 
 +\frac{(1 - \l^2) \sinh\at \cosh\at}{\tilde{\D}} \G^2
 \\[5pt]
 &\left. + \frac{(1 - \l^2) \cos\a \cosh\at}{\sqrt{\D \tilde{\D}}} (i\s_2)\G^2
 -\frac{(1 - \l^2) \sin\a \sinh\at}{\sqrt{\D \tilde{\D}}} (i\s_2)\G^3 \right)\e 
  \\[5pt]
  &=-\frac{\mu(1-\lambda)}{2}\left(\frac{\sinh\tilde\alpha}{\sqrt{\tilde\Delta}}+\frac{\cos\alpha}{\sqrt{\Delta}}i\sigma_2\right)
  (\sigma_1P_{\tilde\alpha}-\sigma_3 P_\alpha)\epsilon\ ,
 \end{aligned}
\end{equation}
 where $\m$ is defined in \eqn{mmm} and vanishes for $\l=0$. 
For $\l\neq 0$ setting \eqn{fhgu1}  to zero is solved by imposing the extra projection
\begin{equation}
 \label{ProjectorLambdane0}
 \sigma_1P_{\tilde\alpha}=\sigma_3 P_\alpha\ ,
\end{equation}
where we have defined
\begin{equation}
 \begin{aligned}
  & P_{\at} = \frac{(1 + \l) \cosh\at\, \s_1\G^2 - (1 - \l) \sinh\at \big( i \s_2 \big)\G^{012} }{\sqrt{\tilde{\D}}} \, ,
 \\[5pt]
  & P_\a = \frac{(1 + \l) \sin\a\, \s_1\G^3 + (1 - \l) \cos\a \big( i \s_2 \big)\G^{345} }{\sqrt{\D}} \, .
 \end{aligned}
\end{equation}
The matrices $P_{\at}$ and $P_\a$ satisfy  the properties
\begin{equation}
\begin{split}
&
P_{\a}^2=P_{\tilde\a}^2=1\,,\qquad (i\s_2) P_{\tilde\a}=-P_{-\tilde\a}(i\s_2)\  ,
\\
&
P_\a(i\s_2)=(i\s_2)P_{-\a}\, ,\qquad
P_{\a}P_{-\tilde\a}=P_{\tilde\a}P_{-\a}\ .
\end{split}
\end{equation}
Using the above, one can also show
\begin{equation}
\left( P_\a \big( i \s_2 \right) P_{\at} \big)^2= \left( P_{\at} \big( i \s_2 \big) P_{\a} \right)^2=1\,.
\end{equation}

Shifting to the gravitino variation \eqref{SUSYvariations}, we provide simplified expressions for its components after imposing the type-IIB chirality condition \eqref{chiral} and the projections \eqref{ProjectorLambda0} and \eqref{ProjectorLambdane0}. In particular, the ones that are not solved trivially are:

\noindent \textbf{The $\tilde\a$ component}
\begin{equation}
\label{grav1}
\partial_{\tilde\alpha}\e+\frac{1}{2}\frac{1-\l^2}{\tilde\D}\s_3\G^{01}\e=0\ .
\end{equation}

\noindent \textbf{The $\tilde\b$ component}
\begin{equation}
\label{grav2}
 \partial_{\tilde\beta}\e-\frac{1}{2}\s_3\G^{02}\frac{1}{\sqrt{\tilde\D}}\left((1+\l)\cosh\tilde\a+(1-\l)\sinh\tilde\a\s_3\G^{01}\right)\e=0\ .
 \end{equation}

\noindent \textbf{The $\tilde\g$ component}
\begin{equation}
\label{grav3}
 \partial_{\tilde\g}\e-\frac12\left(\sinh\tilde\beta\G^{01}-\cosh\tilde\beta(i\s_2)\G^1P_{\tilde\a}\right)\e=0\ .
 \end{equation}

\noindent \textbf{The $\a$ component}
\begin{equation}
\partial_\a\e+\frac12\frac{1-\l^2}{\D}\s_3\G^{45}\e=0\ .
\end{equation}

\noindent \textbf{The $\b$ component}
\begin{equation}
\partial_\b\e-\frac12\G^{34}\frac{1}{\sqrt{\D}}\left((1-\l)\cos\alpha+(1+\l)\sin\alpha\s_3\G^{45}\right)\e=0\ .
\end{equation}

\noindent \textbf{The $\g$ component}
\begin{equation}
\partial_\g\e-\frac12\left(\cos\b\G^{45}+\sin\b(i\s_2)\G^4P_\a\right)\e=0\ .
\end{equation}
Obviously the first three can be integrated independently from the last ones. Starting with the $\at$ component we see that it can be integrated straightforwardly giving
\begin{equation}
\label{res1}
\e=\Omega_{\tilde\a}\e_1\,,\quad \Omega_{\tilde\a}=\exp\left(-\frac12\tanh^{-1}\left(\frac{1-\l}{1+\l}\tanh\tilde\a\right)\s_3\G^{01}\right) \,.
\end{equation}
Here $\e_1$ is a spinor which depends on $(\bt, \gt , \a , \b , \g)$.
Then the equation in the $\bt$-direction \eqref{grav2} reduces significantly if we make use of the latter result. Namely, we find that
\begin{equation}
\partial_{\tilde\b}\e_1-\frac12\s_3\G^{02}\e_1=0 \, .
\end{equation}
Again, a simple integration gives
\begin{equation}
\label{res2}
\e_1=\Omega_{\tilde\b}\e_2\,,\quad \Omega_{\tilde\beta}=\exp\left(\frac\bt2\s_3\G^{02}\right)\e_2 \, ,
\end{equation}
where now $\e_2$ is a spinor that depends on $(\gt , \a , \b , \g)$. If we now combine \eqref{res1} and \eqref{res2} we find that the equation in the 
$\gt$-direction \eqref{grav3} condenses to
\begin{equation}
\partial_{\tilde\g}\e_2+\frac12\s_3\G^{12}\e_2=0 \, .
\end{equation}
Once more, this can also be easily integrated to
\begin{equation}
\label{res3}
\e_2=\Omega_{\tilde\g}\e_3\,,\quad \Omega_{\tilde\g}=\exp\left(-\frac\gt2\s_3\G^{12}\right)\e_3 \, ,
\end{equation}
with $\e_3$ being a spinor that depends on $(\a , \b , \g)$.
The other three equations of the gravitino variation can be solved in the same fashion. This process confines the spinor $\e_3$ to the form
\begin{equation}
\label{res4}
\e_3=\Omega_a\Omega_\b\Omega_\g\eta\,,
\end{equation}
where $\eta$ is a constant spinor and
\begin{equation}
\label{res5}
\begin{split}
&\Omega_\a=\exp\left(-\frac12\tan^{-1}\left(\frac{1+\l}{1-\l}\tan\a\right)\s_3\G^{45}\right)\,,\\
&\Omega_\b=\exp\left({\frac\b2\G^{34}}\right)\,,\quad \Omega_\g=\exp\left({\frac\g2\G^{45}}\right) \, .
\end{split}
\end{equation}
To summarize, combining \eqref{res1}, \eqref{res2}, \eqref{res3}, \eqref{res4} and \eqref{res5} we find that the Killing spinor $\e$ is
\begin{equation}
 \label{SpinorFinal}
 \e = \Om_{\at }\Om_{\bt} \Om_{\gt} \Om_{\a} \Om_\b \Om_\g\eta \, ,
\end{equation}
which is precisely \eqref{KillingSpinorLambda}.

Before we conclude, we need to show that the projections \eqref{ProjectorLambda0} and \eqref{ProjectorLambdane0} reduce to algebraic constraints on $\eta$ with no coordinate dependence. Starting with \eqref{ProjectorLambda0}, we notice that the matrix $\G^{012345}$ freely passes through the $\Om$ matrices in \eqref{SpinorFinal}. Therefore, equation \eqref{ProjectorLambda0} reduces to \eqref{ProjectorLambda0eta}. The case \eqref{ProjectorLambdane0} needs more care. For convenience we write it as
\begin{equation}
 P_{\a} (i \s_2) P_{\at} \e = \e \, , \qquad P_{\a} (i \s_2) P_{\at} = \Om^2_{\a} \Om^2_{\at} \s_1 \G^{2345} \,,
\end{equation}
where we made use of the relations
\be
\Omega_{\tilde\a}^2=P_{\tilde\a}\s_1\G^2\,,\quad \Omega^2_\a=P_\a(i\s_2)\G^{345}\,.
\ee

Using the commutation properties for the $\G^a$ and $\s_i$ matrices it is easy to see that the above formula implies
\begin{equation}
 \s_1 \, \G^{2345} \eta = \eta \,,
\end{equation}
which reduces to \eqref{ProjectorLambdane0eta} if we combine it with \eqref{ProjectorLambda0eta}.

\subsection{Supersymmetry variations for the pp background}
\label{ppsusy} 

We now consider the analysis of the supersymmetry for the background of the Section \ref{section.pp}. Starting with the dilatino equation \eqref{SUSYvariations} it is easy to see that it takes the simple form\footnote{
We use the frame $(e^+,e^-,e^{1\ldots8})$
\begin{align}
  & e^{+} = dv - \frac{1}{2} \bigg[\Big( \frac{1 - \l}{1 + \l} \Big)^4 \big( z^2_1 + z^2_2 \big) +z^2_3 + z^2_4 \biggr] du \, , \quad
  e^{-} = du \, , \quad 
  e^0 = \frac{1}{\sqrt{2}} \big( e^{+} - e^{-} \big) \, , \quad
  e^9 = \frac{1}{\sqrt{2}} \big( e^{+} + e^{-} \big) \, ,
  \nonumber\\[5pt]
  & e^1 = dz_1 \, , \quad e^2 = dz_2 \, , \quad e^3 = dz_3 \, , \quad e^4 = dz_4 \, , \quad
  e^5 = dx_1 \, , \quad e^6 = dx_2 \, , \quad e^7 = dx_3 \, , \quad e^8 = dx_4 \, .
  \label{ppFrame}
 \end{align}
}
\begin{equation}
\label{pp.dilatino}
  \d\l= \frac{1}{(1+\l)^2}\s_3\G^{13} \left( 1 + \l^2  - 2\l (i \s_2) \G^{34} \right) \G^{-} \big( 1 - \G^{1234} \big) \e = 0 \, .
\end{equation}
From the above one might suspect that when $\l \ne 0,1$ the pp-wave preserves $24$ supercharges
\be
\label{susyl0}
\G^{-} \big( 1 - \G^{1234} \big) \e = 0 \,,
\ee
while for $\l = 1$ supersymmetry it enhances to $28$ supercharges
\be
 \big( 1-  (i \s_2) \G^{34} \big) \G^{-} \big( 1 - \G^{1234} \big) \e = 0 \, .
\ee
However, one has also to analyze the gravitino equation \eqref{SUSYvariations}. 

Working out the gravitino components $\d\psi_+$ and $\d\psi_i \, (i = 1 , \ldots , 8)$, we find
\begin{equation}
 \partial_v \e = \partial_i \partial_j \e = 0 \qquad \forall i, j= 1 , \ldots , 8\, .
\end{equation}
This suggests that the Killing spinor $\e$ must be independent of the coordinate $v$ and linear in $x_i \, , z_i \, (i = 1 , \ldots , 4)$. In other words, it takes the form of \eqref{spinorexpression}, where $\chi$ is a spinor that depends on $u$ and the matrix $\Om$ is given by
\begin{equation}
\label{Omega.matrix}
 \begin{aligned}
  \Om = & - \frac{1}{2} \frac{1 + \l^2}{(1 + \l)^2} \s_3 \Big( \G^3 z_1 - \G^4 z_2 - \G^1 z_3 + \G^2 z_4 \Big)
  \\[5pt]
  & - \frac{1}{2} \frac{\l}{(1 + \l)^2} \s_1 \Big( \G^4 z_1 + \G^3 z_2 - \G^2 z_3 - \G^1 z_4 \Big)\left(1+\G^{1234}\right)
  \\[5pt]
  & +\frac12 \frac{\l}{(1 + \l)^2} \s_1 \big( 1 - \G^{1234} \big) \G^{23} \Big( \G^5 x_1 + \G^6 x_2 + \G^7 x_3 + \G^8 x_4 \Big) \, .
 \end{aligned}
\end{equation}
The spinor $\chi(u)$ is determined by analysing the component $\d\psi_-$. Using the form \eqref{spinorexpression} of the Killing spinor, we find that $\d\psi_-$ reduces to
\begin{equation}
 \label{deltapsim}
 \begin{aligned}
  &(1+ \G^{-} \Om) \, \partial_u \chi + \frac{1}{4} \partial_{z_i} \cH \, \G^{- i} \chi - \frac{1}{2} \frac{1 + \l^2}{(1 + \l)^2} \s_3 \big( \G^{13} - \G^{24} \big) \big( 1 + \G^{-} \Om  \big)\chi
  \\[5pt]
  & - \frac{1}{2} \frac{\l}{(1 + \l)^2} \s_1 \G^{-+} \big( \G^{14} + \G^{23} \big) \big( 1 + \G^{-} \Om  \big) \chi= 0 \, ,
 \end{aligned}
\end{equation}
where
\begin{equation}
 \cH = - \Big( \frac{1 - \l}{1 + \l} \Big)^4 \big( z^2_1 + z^2_2 \big) - z^2_3 - z^2_4 \, .
\end{equation}
Due to the linear dependence of the above on $z_i \, , x_i \, (i = 1, \ldots,4)$, we can derive a set of conditions for the spinor $\chi(u)$. This can be done by differentiating \eqref{deltapsim} with respect to the aforementioned coordinates.

Starting with the piece of \eqref{deltapsim} that is independent of $z_i \, , x_i \, (i = 1, \ldots,4)$ we obtain the following differential equation for $\chi(u)$
\begin{equation}
\label{Killing.u}
 \partial_u \chi - \frac{1}{2} \frac{1 + \l^2}{(1 + \l)^2} \s_3 \big( \G^{13} - \G^{24} \big) \chi - \frac{1}{2} \frac{\l}{(1 + \l)^2} \s_1 \G^{-+} \big( \G^{14} + \G^{23} \big) \chi = 0 \, .
\end{equation}
Acting with $\partial_{z_i} \, , \partial_{x_i} \, (i = 1 , \ldots , 4)$ on \eqref{deltapsim} and using \eqref{Killing.u} gives the conditions
\be
\label{pp.conditions}
\begin{split}
&\l\G^-\left(2(1+\l^2)-\l+\l\G^{1234}+(1+\l^2)(i\s_2)(\G^{34}-\G^{12})\right)\chi=0\,,\\
&\l\G^-\left(-2(1+\l^2)-\l+\l\G^{1234}-(1+\l^2)(i\s_2)(\G^{34}-\G^{12})\right)\chi=0\,,\\
&\l\G^-\left(1-\G^{1234}\right)\chi=0 \, .
\end{split}
\ee
The first is obtained from the action of $\partial_{z_i} \, (i = 1, 2)$, the second from $\partial_{z_i} \, (i = 3, 4)$ and the last from $\partial_{x_i} \, (i = 1 , \ldots , 4)$. Together with the above, one should also take into account the constraint coming from the dilatino equation \eqref{pp.dilatino}. Using \eqref{spinorexpression} this translates to
\begin{equation}
 \label{dilatinochi}
 \Big( 1 + \l^2  - 2\l (i \s_2) \G^{34} \Big) \G^{-} \big( 1 - \G^{1234} \big) \chi = 0 \, .
\end{equation}

\no
We conclude by observing that the combinations of Pauli and $\G$ matrices appearing in \eqref{Killing.u} commute with each other. Therefore, a straightforward integration  results to \eqref{spinor.sol}.

\end{document}